\documentstyle[12pt]{article}
\begin{document}
\begin{center}
{\Large \bf
Phase structure and confinement properties of noncompact
gauge theories	\\
II. $Z(N)$ Wilson loop and effective noncompact model}\\
\vspace{1cm}
{\large O.A.~Borisenko, V.K.~Petrov, G.M.~Zinovjev
\footnote{email: gezin@gluk.apc.org}}\\
{\large \it
N.N.Bogolyubov Institute for Theoretical Physics, National Academy
of Sciences of Ukraine, 252143 Kiev, Ukraine}\\
\vspace{1cm}
{\large J.~Boh\'a\v cik,
\footnote{email: bohacik@savba.savba.sk}}\\
{\large \it
Institute of Physics, Slovak Academy of Sciences,
84228 Bratislava, Slovakia}\\

\end{center}
\vspace{.5cm}

\begin{abstract}
An approach to studying lattice gauge models in the weak coupling
region is proposed.  Conceptually, it is based on the crucial role of
the original $Z(N)$ symmetry and the invariant gauge group measure. As
an example, we calculate an effective model from the compact Wilson
formulation of the $SU(2)$  gauge theory in $d=3D3$ dimensions at zero
temperature. Confining properties and phase structure of the effective
model are studied in details.
\end{abstract}

\newpage

\section{Introduction}
In the previous article \cite{ncmpt1} we presented a brief review of
noncompact lattice models and considered several
examples aiming to clarify some aspects of the confinement problem
in the noncompact formulation of lattice theories.
Our conclusion was that one may construct a noncompact model
which confines in the same way as compact model does if we take properly
into account the contribution of the invariant measure and the centre
subgroup of the original $SU(N)$ symmetry of the compact model.
We are going to demonstrate here how to calculate a noncompact
model from the Wilson compact formulation including both these conditions.
It seems to be worthwhile to start our discussion with a brief summary of
earlier investigations \cite{ncmpt1}.

We calculated the $SU(2)$ partition function of the noncompact model
at finite temperature approximating the initial action by its
chromoelectric part (time-like plaquettes) and compared it with
known expression for the $SU(2)$ partition function of the compact
Wilson model in the same approach. We have found out that these two
models coincide at high temperatures only and the noncompact version
does not exhibit the confining features. The same result was obtained
by avoiding an expansion of Polyakov loops around the unit matrix and
preserving the compact integration over $A_0$ gauge potential.
We get convinced by this example that compact Wilson models and
noncompact lattice Yang-Mills theories might belong to different
universality classes.

As the next step, we followed a proposition of Ref. \cite{pol} and
considered the sine-Gordon model with a $Z(2)$ invariant potential.
This potential had to reproduce a contribution of the invariant group
measure for $A_0$ gauge potential in the noncompact formulation.
However, the model occurred to break  the global $Z(2)$ symmetry at all
values of the coupling constants in $d>2$ dimensions \cite{gopfert}.
This fact seems to be in contradiction with the main idea of \cite{pol}
since the invariant measure was introduced to preserve the $Z(2)$ symmetry
of the vacuum. The behaviour of some important correlation functions
in the sine-Gordon model resembles rather that of $U(1)$ Villain lattice
model investigated in \cite{gopfert} and the sine-Gordon model is
an effective model of the lattice abelian theory with the Villain action.

We concluded then that in order to reproduce
the specific features of the $SU(2)$ Wilson theory we have to preserve
not only the global centre symmetry but the local centre symmetry as well.
In the absence of the local $Z(2)$ symmetry the global one can be
spontaneously broken what is the case for the sine-Gordon model
whereas there should be no such a breakdown in zero temperature $SU(2)$
model. Actually, the invariance of the theory with respect to
global $Z(2)$ arises only from the potential term in the sine-Gordon model
whereas the kinetic term is invariant under global $U(1)$ transformations.
Meanwhile, compact Wilson action itself is invariant both under $Z(2)$
global and under $Z(2)$ local transformations. It is a very essential point
in the $Z(N)$ mechanism of confinement that $SU(N)$ gauge group is broken
up to its local $Z(N)$ subgroup \cite{mack,mack2}.
Hence, accepting the idea of Ref.\cite{pol} that
one should simulate the invariant measure in a form of a local $Z(N)$
invariant potential we believe that an effective noncompact action should
be itself invariant under $Z(N)$ local transformations but not only under
global ones. To do so, it is not sufficient to introduce invariant measure
into the effective action. A modification of the model \cite{pol}
which respects {\it local} $Z(N)$ symmetry will be presented in this articl=
e.
The second important observation which can be extracted from
these discussions is the conclusion that we are not allowed
to neglect dynamics of space gauge potentials. At least, we have to
reproduce the contribution of $Z(N)$ configurations contained
in $U_n(x)$ matrices.

Another hint supporting these ideas comes from the discussion of a version
of the $XY$ model which can be an effective model of $SU(2)$ gauge theory
at finite temperature \cite{ncmpt1,su2gl} in the strong coupling region.
The corresponding effective action [see Eq.(56) of \cite{ncmpt1}]
has only a global $Z(2)$ invariance
which can be spontaneously broken at high temperature. However, the
corresponding effective action at zero temperature has also local
$Z(2)$ symmetry (see next section for details) and this is, according to
Elitzur's theorem, the reason of the absence of the phase transition at zero
temperature in nonabelian gauge theories. If we set space gauge potentials
to be equal to zero (considering their dynamics as being inessential
for the confinement, as has been proposed in \cite{pol}) we get either the
compact $XY$ model or the noncompact sine-Gordon model. In both cases
the spontaneous breaking of $Z(2)$ symmetry takes place (actually,
the resulting theory is invariant under global $U(1)$ transformations and
just this symmetry is completely broken as it happens in zero temperature
$U(1)$ lattice model where deconfinement of electrons takes place).
To avoid this breakdown it is necessary to perform a summation over $Z(2)$
configurations contained in space gauge fields. In such a way we shall come
to the theory with a local $Z(2)$ symmetry. This is what will be done in the
next section. The corresponding noncompact theory could then have
a form of the sine-Gordon model with local $Z(2)$. Doing so, we can
construct a  noncompact weak coupling limit of the compact model
which exhibits confinement as compact theory does.
We showed (this was shortly discussed in \cite{l94}, too) how to
define a noncompact model with necessary confining properties,
but we did not give a definition of the correlation functions
of Polyakov loops and Wilson loops. We will fill this gap in the present
paper.

Thus, our basic observation is if we are willing to construct a noncompact
theory confining in the same way as compact one, we have to 1) perform
the summation over $Z(N)$ variables in the compact model and then to take
a limit $g \rightarrow 0$ expanding gauge field matrices around unit matrix;
2) include invariant measure in a form of a local $Z(2)$ invariant potentia=
l.
We need to be very careful using the expansion of gauge
field matrices around unit ones because we can miss relevant minima
of an effective action which we get after summation over $Z(N)$
configurations. We have demonstrated this point in the previous article on
a simple example of the noncompact model with compact integration over
$A_0$ gauge potential. The expansion around all minima is required
to come to the reliable model in the weak coupling region.

The outline of the paper is as follows.

In section 2 we discuss in detail a general strategy of our investigation
of the Wilson lattice theory in the weak coupling region. We derive
a representation for the partition function of the 3-dimensional $SU(2)$
gauge theory performing the summation over $Z(2)$ configurations.
In section 3 we define the $XY$ model with a local $Z(2)$ symmetry
and study some of its properties. Here, we also present our
effective noncompact model in the weak coupling region.
The Wilson loops are defined in our model in section 4. We
calculate $Z(2)$ Wilson loop in some simplest cases and show the
finiteness of the string tension in the limit of vanishing bare coupling
constant (continuum limit). We terminate with section 5 as a Summary
discussing essential points of our approach, confinement mechanism
in gauge theories as well as some still unsolved problems.

\section{General representation for the partition function}

We are going to accomplish the above programme taking as an example
$SU(2)$ lattice gauge theory (LGT) in $d=3D3$ dimensions.
The corresponding partition function has the form
\cite{wilson}
\begin{equation}
Z=3D\int D\mu (U)  \exp \left [ \lambda \sum_{p} \Omega_{\partial p}(U)
\right ],
\label{1}
\end{equation}
\noindent
where $\Omega$ is a character of the fundamental representation of $SU(2)$
gauge group, $D \mu (U)$ is the invariant integration measure and
$\lambda =3D \frac{2}{g^{2}}$. We are interested in a reliable expansion
for the gauge matrices in the weak coupling region. It is obvious, that
configurations dominating the partition function (\ref{1}) in this
limit are those for which $\Omega_{\partial p}(U)$ is close to unity.
It does not follow, however, that a single gauge matrix $U_{\mu}(x)$ is
close to unity. The naive expansion of $U_{\mu}(x)$ around unit matrix leads
to the usual perturbative expansion implying the loss of confinement.
Nevertheless, we may use this expansion if we choose a proper gauge which
makes all matrices $U_{\mu}(x)$ as close to unity as possible. This
so-called minimal Landau gauge was studied in details in \cite{zw1,zw2}.
An effective action calculated in the thermodynamical limit differs from
the Faddeev-Popov action by inclusion of a new term which preserves,
however, renormalizability and asymptotic freedom. The corresponding weak
coupling expansion differs from the Faddeev-Popov perturbation theory
by terms which are finite in every order and are small only at high
energies \cite{zw1}. Under some plausible assumptions an area law for
the Wilson loop in the newly developed theory has been proved \cite{zw2}.
Though we are fully aware of the success of this approach in
the study of the weak coupling region of LGT we would like to develop
a method which, on our view, could clarify in some aspects the
role of $Z(N)$ symmetry and $Z(N)$ excitations in confinement
and could be useful in an investigation of a mechanism of this phenomenon.

We rewrite (\ref{1}) using a representation
\begin{equation}
\Omega_{\partial p}(U) =3D Z_{\partial p} \frac{1}{2} \mbox{Tr}
\bar{U}_{\partial p},
\label{2}
\end{equation}
\noindent
where $\bar{U}_{\partial p} \in SU(2)/Z(2) \sim SO(3)$ and $Z_{\partial p}$
is a product of $Z(2)$ elements along the minimal plaquette.
For the invariant measure we have
\begin{equation}
D \mu (U) =3D \frac{1}{2} \sum_{z=3D\pm 1} D \mu (\bar{U}),
\label{3}
\end{equation}
\noindent
where $D \mu (\bar{U})$ is an invariant measure on $SO(3)$ group.
Let us recall, that invariant measure on $SU(N)/Z(N)$ group coincides with
$SU(N)$ measure up to the restriction
\begin{equation}
-\frac{\pi}{N} \leq arg[TrU] \leq \frac{\pi}{N}.
\label{o3meas}
\end{equation}
\noindent
Setting $\bar{U}_{\mu}(x)=3DI$ everywhere, we get a $Z(2)$ gauge model which
confines static charges in the strong coupling region. $Z(N)$ gauge
model can be viewed as a classical solution of the $SU(N)$ Wilson model
\begin{equation}
U_{\mu}(x) =3D V_{x}Z_{\mu}(x)V_{x+\mu}^+ ,
\label{4}
\end{equation}
\noindent
where $V_x$ is an arbitrary $SU(N)$ matrix \cite{yon}. The space-time
structure of this solution is a two-dimensional closed surface and in the
continuum limit it corresponds to singular gauge transformations.
It should be emphasized that this solution,
representing $Z(N)$ excitations, is one of the few exact topological
solutions of the Wilson model which could, in many cases, provide
the same minimum of the Wilson action as the trivial one
$U_{\mu}(x) =3D V_{x}V_{x+\mu}^+$. In our case this minimum is achieved
when the product of $Z(N)$ elements along  minimal plaquettes
equals $1$. However, the $Z(N)$ gauge model undergoes a phase transition
to the deconfinement phase in the weak coupling region. It was proposed in
\cite{yon} to calculate quantum corrections for $Z(N)$ excitations
coming from the following expansion for $SO(3)$ gauge matrices
\begin{equation}
\bar{U}_{\partial p} \approx 1 - F^{2}_{\mu \nu},
\label{5}
\end{equation}
\noindent
where $F_{\mu \nu}(A)$ is the strength tensor,
aiming to avoid the phase transition from the $SU(N)$ LGT (under assumption,
of course, that $Z(N)$ configurations play a central role in a formation of
the confining forces). As a result, we have $Z(N)$ gauge spin system connec=
ted
to noncompact Yang-Mills potentials. It has been found, however, that the
phase transition still exists, although the critical coupling moves to a
smaller value. Thus, the string tension is vanishing in the weak coupling
region. We have to conclude that the expansion (\ref{5}) is not suitable
in this region because no phase transition has been found in $SU(2)$ and
$SU(3)$ gauge theories
(it is possible that a roughening phase transition can happen since
it is of an infinite order and does not affect the string tension) and leads
to the false vacuum in the weak coupling region. Indeed, when the product
of $Z(2)$ elements along a plaquette equals $1$ then the expansion (\ref{5})
provides the global minimum of the action (if we do not fix a gauge).
However, this expansion does not provide the same minimum when
$Z_{\partial p} =3D -1$. We think that this is the main reason why
$Z(N)$ global symmetry occured to be broken when $g \rightarrow 0$
(we shall specify mentioned symmetry below). Our approach in investigation
of the role of $Z(2)$ configurations in the region of small coupling
consists in the following. We expand the space gauge matrices around the un=
it
matrix similarly to \cite{yon}
\begin{equation}
\bar{U}_{n}(x) =3D 1 + iagA_n(x).
\label{6}
\end{equation}
\noindent
However, unlike \cite{yon} this expansion can be made more rigorous if we
fix a gauge which makes all the matrices $\bar{U}_{n}(x)$ as close to unity
as possible. Since $Z(2)$ transformation under consideration
acts on ${U}_0$ gauge matrices, we choose not to expand these matrices
taking them in a diagonal form  but not fixing a static gauge
(this is, of course, an approximation which leads, however, to the same
permanent features discussed below as an exact treatment \cite{sur}).
This procedure results in a kind of the $XY$ model for the
$A_0$ gauge potential\footnote{In an exact procedure we would get
a kind of $O(2)*O(2)$ model for $A_0$.}.

Thus, our starting point is the partition function
\begin{eqnarray}
Z =3D \sum_{z} \int \prod_{x,n} DA_{n}(x) \prod_{x} D\bar{U}_{0}(x)
\exp [ \frac{\lambda}{2} \sum_{\bar{p}} Z(\partial \bar{p})
\mbox{Tr}(1 - \frac{a^4 g^2}{2} F^{2}_{nm}(x)) +  S_d  +  S_{g.f.}
\nonumber  \\
+ \frac{\lambda}{2} \sum_{p_{0}} Z(\partial p_0) \mbox{Tr}
\bar{U}_0(x)(1 + iagA_n(x))\bar{U}_0^+(x+n)(1 - iagA_n(x+m)) ],
\label{7}
\end{eqnarray}
\noindent
where we have omitted an irrelevant constant and introduced the following
notations. $F^{2}_{nm}(x)$ is the Yang-Mills strength tensor, $S_{g.f.}$
is the gauge fixing term and $S_d$ is the Faddeev-Popov determinant
(here and futher Latin (Greek) indices mean two(three)-dimensional vectors,
bars mean two-dimensional quantities). $\bar{p}$ ($p_0$) is the
space(time)-like plaquette. The gauge fixing term could be taken, for
example, as in \cite{zw1} where we have to substitute $SU(2)/Z(2)$
matrices instead of $SU(2)$ ones.
$\int \prod_{x,n} DA_{n}(x)$ is an integral over gauge potentials calculated
over all noncompact region, while $D\bar{U}_{0}(x)$  means the invariant
measure on $SO(3)$ group. We suppose, however, that the dynamics of space
gauge potentials is not important in achieving an area law for the time-spa=
ce
Wilson loop even in the continuum limit\footnote{The only difference
in this respect from \cite{pol} is that we keep centre elements of
space-gauge matrices}.
In this approach $S_{g.f.}$ and
$S_d$ might be excluded from the consideration as we have fixed the gauge
only for $SO(3)$ part of the gauge group and thereby all these terms do not
depend on $Z(2)$ variables. In this approach we have
the following expression for the partition function:
\begin{equation}
Z =3D \sum_{z_{\mu}(x)} \int \prod_{x} D\bar{U}_{0}(x)
\exp \left [\lambda \sum_{\bar{p}} Z(\partial \bar{p}) +
\lambda \sum_{p_{0}} Z(\partial p_0)  \cos (\phi_{t}(\bar{x})
- \phi_{t}(\bar{x}+n)) \right ],
\label{9}
\end{equation}
\noindent
where we have denoted $\phi_{t}(\bar{x}) =3D agA_0(x)$.
Let us discuss briefly what shall we get if we do not set $A_n(x) =3D 0$?
We can follow \cite{yon} and consider the quadratic fluctuations around
$Z(2)$ space configurations fixing the Feynman gauge, i.e.
\begin{equation}
S_{g.f.} =3D -\frac{1}{2} \sum_{x,n} [\Delta_n A^c_{n}(x)]^2,
\label{10}
\end{equation}
\noindent
where $\Delta_n$ is the difference operator on the lattice.
This results in the gaussian path integral over space gauge potentials
$A_n(x)$ calculating which we come to a bosonic determinant in the
background $Z(2)$ and $A_0(x)$ gauge fields. Since in what follows
we shall study (\ref{7}) in the approach $A_n(x) =3D 0$, we adduce here
only qualitative arguments why such integration cannot change
the confinement picture which follows from the partition function
(\ref{9}). The main effects of the discussed integration appear to be
the following. First of them was pointed out in \cite{yon}.
Expanding the determinant in powers of the fluctuations
$Z(\partial \bar{p}) - 1$, one finds that the first term generates
the Wilson action leading to the substitution
\begin{equation}
\lambda \rightarrow \lambda - \frac{N_{c}^{2}-1}{4}
\label{11}
\end{equation}
\noindent
in (\ref{9}). Obviously, it is not difficult to take into account
this contribution. The second power in $Z(\partial \bar{p}) - 1$,
as has been shown in \cite{yon}, leads to an increasing of the disorder
of the system and, as such, can only enhance the string tension and
lower the critical coupling. It is very complicated to estimate reliably
the contribution of higher order corrections.
Another effect coming from time-like plaquettes is the generation
of a new interaction between $Z(2)$ excitations and $A_0(x)$ gauge
potential of the kind
\begin{equation}
\sum_{p_{0}} Z(\partial p_0)  \cos (\phi_{t}(\bar{x})
+ \phi_{t}(\bar{x}+n)).
\label{12}
\end{equation}
\noindent
Being added to the action in (\ref{9}) this term reduces the original
symmetry of the action to $Z(2)$ symmetry. Let us explain this in
more detail. The original action (\ref{1}) as well as the action in
(\ref{7}) is invariant under global $Z(2)$ transformations
(diagonal gauge for $U_0$) $\phi (x) \rightarrow \phi (x) \pm \pi$
whereas it could seem that the action in (\ref{9}) is invariant
under global $U(1)$ transformations
$\phi (x) \rightarrow \phi (x) \pm const$. The integration  over
space gauge potentials recovers the original symmetry of the action.
It does not follow, however, that the theory defined in (\ref{9}) is
invariant under more general $U(1)$ transformations since the group
measure in the integrand of (\ref{9}) is invariant only under $Z(2)$.
In this paper we neglect the contributions of the type (\ref{12})
because as we have explained the symmetries of both actions are actually
the same. Moreover, it has been discussed in \cite{ncmpt1} that the terms
in the lattice action like $\cos (\phi (x) + \phi (x+n))$ correspond
in the continuum limit to the contribution of the local potentials
up to the order $O(g)$ which we do not take into account here.
Concerning the contribution of the local potential we assume that
all such corrections are incorporated in the invariant measure
which we simulate as a local $Z(2)$ invariant potential of the
sine-Gordon type.

Let us turn now to the expression (\ref{9}) for the partition function.
We are going to perform the summation over $Z(2)$ variables. One
possible way is to rewrite this sum as a sum over all closed surfaces
in $d$-dimensional space. To do that, we present (\ref{9}) as
\begin{eqnarray}
Z =3D \sum_{z_{\mu}(x)} \int \prod_{x} D\bar{U}_{0}(x)
\prod_{\bar{p}} \left [ \cosh \lambda + \sinh \lambda  Z(\partial \bar{p})
\right ]   \nonumber   \\
\prod_{p_{0}} \left [ \cosh (\lambda \cos \Delta \phi )
+  \sinh (\lambda \cos \Delta \phi) Z(\partial p_0) \right ].
\label{13}
\end{eqnarray}
\noindent
The notation $\Delta \phi =3D \phi_{t}(\bar{x}) - \phi_{t}(\bar{x}+n)$
has been used. Summing over $z_{\mu}(x)$ we notice that the only plaquette
configurations giving a non-zero contribution to $Z$ are those which form
closed surfaces  (every link must enter even number of times in the product
over all plaquettes). As a result we have a sum over all possible products
of all possible surfaces on the lattice. The surfaces which enter the same
product have no plaquettes in common. Let $\Omega$ be an arbitrary closed
2-surface constructed from $\mid \Omega \mid$ plaquettes, $N_{\bar{p}}$
($N_{p_{0}}$) - a number of space(time)-like plaquettes and $N_l$ - a
number of links. $N_p$ is a common number of plaquettes on the lattice.
We put down the result in the form, recovering all constants
\begin{equation}
Z =3D e^{-\lambda N_p} (\cosh \lambda )^{N_{\bar{p}}} 2^{N_l}
\sum_{\Omega} (\tanh \lambda )^{\mid \Omega_{\bar{p}} \mid}
\sum_{\sigma =3D \pm 1} (\prod_{p_0 \in \Omega} \sigma ) Z_{\Omega}(\sigma).
\label{14}
\end{equation}
\noindent
We have introduced here new time-like plaquette's variable $\sigma$ for
further convenience. Since the action $\sum_{p_0}\cos \Delta \phi$
is diagonal in time indices, we can consider $\sigma (p_0) =3D
\sigma_{n}(t)$ as link variable and then define
\begin{equation}
Z_{\Omega}(\sigma) =3D  \prod_{t=3D1}^{N_t} Z^{0}_{\Omega}(\sigma_{n}(t)) ,
\label{15}
\end{equation}
\noindent
\begin{equation}
Z^{0}_{\Omega}(\sigma_{n}(t)) =3D 2^{-N_p} \int \prod_{x} D\bar{U}_{0}(x)
\exp [ \lambda \sum_{p_0} \sigma \cos \Delta \phi ].
\label{16}
\end{equation}
\noindent
The summation over $\sigma$ can be easily performed to produce
\begin{equation}
\sum_{\sigma =3D \pm 1} \prod_{p_0 \in \Omega} \sigma
Z^{0}_{\Omega}(\sigma ) =3D
\int \prod_{x} D\bar{U}_{0}(x) \prod_{p_0} \cosh (\lambda \cos \Delta \phi )
\prod_{p_0 \in \Omega} \tanh (\lambda \cos \Delta \phi ).
\label{17}
\end{equation}
\noindent
In what follows we use both representation (\ref{14})-(\ref{16}) and
(\ref{14}),(\ref{17}) for the partition function. It is clear from the
described procedure that this method can be directly applied for the full
action in (\ref{7}). Two additional terms, supported on 2-closed surfaces,
would appear in this case: the Yang-Mills term $F_{nm}^{2}(x)$ and a term
describing interactions between noncompact Yang-Mills potentials and
the compact $\bar{U}_0$ gauge field. The form of the latter can be found
in \cite{ncmpt1} (terms $S_1$ and $S_2$ from the section 4).

Completing this section we want to discuss general phase structure
of the model (\ref{14}) and a representation for the invariant measure
$D\bar{U}_{0}$ on $SO(3)$ gauge group. In absence of the interaction
term in time-like plaquettes, i.e. $\cos \Delta \phi =3D 1$, we get the
three-dimensional $Z(2)$ gauge model. On the dual lattice it is equivalent
to the $Z(2)$ Ising model which has a phase of spontaneously broken
$Z(2)$ global symmetry. The corresponding phase transition on the original
lattice corresponds to the deconfinement phase transition. In the weak
coupling region the Wilson loop obeys the perimeter law  behaviour.
The term $\cos \Delta \phi$ might
drastically change this picture. On the dual lattice the leading
configurations contributing to the partition function in the limit
$\lambda \rightarrow \infty$ are those when all dual spins are strongly
disordered (ordered, deconfinement phase on the original lattice).
At small $\lambda$ the main contribution comes from the ordered
configurations of the dual spins (this is the confinement phase on the
original lattice). The term $\cos \Delta \phi$ generates contributions
which negative plaquettes $Z_{\partial p}=3D-1$ turn into positive ones,
thus disodering the $Z(2)$ system on the original lattice even at large
$\lambda$. Our hope is that this competition between the ordering
mechanism of the $Z(2)$ system and disordering mechanism of the $XY$ system
will shift the critical point to infinity, i.e. $g \rightarrow 0$.
A roughening phase transition of the infinite order is still possible
at a finite value of $\lambda$. We shall turn to this principal point
again in section 4.

We will see a hint on another phase transition if we neglect the
space-like plaquettes in (\ref{9}). Summing over $z_{\mu}(x)$ in this
approach one has in the thermodynamical limit on the periodic lattice
\begin{equation}
Z =3D 2^{N_l}  \int \prod_{x} D\bar{U}_{0}(x)
\prod_{p_0} \cosh (\lambda \cos \Delta \phi ).
\label{18}
\end{equation}
\noindent
This model resembles the $XY$ model on the square lattice which undergoes
a phase transition related to the condensation of vortices.
However, in addition to the usual $U(1)$ global symmetry, the model
(\ref{18}) possesses the local $Z(2)$ symmetry (see discussion in the next
section). This property immediately means that any $Z(2)$ nonivariant
observable or correlation function is identically equal to zero.
$Z(2)$ invariant correlation function displays qualitatively different
behaviour in the limits $\lambda \rightarrow 0$ and $\lambda \rightarrow
\infty$ when $D\bar{U}_{0}(x) =3D d\phi (x)$ like in the standard $XY$ mode=
l.
As will be discussed further, such correlation functions are related to the
adjoint sources introduced in the original model (\ref{1}). As well known,
however, adjoint sources are screened in pure $SU(N)$ gauge theories
at large distances \cite{seiler}
in the strong coupling limit and this behaviour is expected
to be valid in the whole range of the coupling constant. We conclude from
this that the $SO(3)$ part of the invariant measure could play an important
role in the phase structure of our effective model. The invariant measure
reduces $U(1)$ global symmetry to the $Z(2)$ global symmetry. The latter
could be spontaneously broken in the $XY$ model. In the presence of the loc=
al
$Z(2)$ it is impossible in force of Elitzur's theorem. Thus, we should expe=
ct
that adjoint sources will not show critical behaviour similar to the
fundamental ones if we take properly into account the invariant measure.
Usually, it is accepted that in the naive limit and in the perturbative
theory we have to replace the invariant measure with the flat measure.
This is not so obvious in the nonperturbative treatment of the weak coupling
limit, and the opposite opinion has been advocated in \cite{pol}.
We accept the following general form for the invariant measure contribution
casted by Ref. \cite{pol}:
\begin{equation}
\int \prod_{x} D\bar{U}_{0}(x) =3D \int_{-\pi /2}^{+\pi /2}
\prod_{x} d \phi_x e^{\mu \sum_x V(\phi_{x})},
\label{19}
\end{equation}
\noindent
where $V(\phi_{x})$ is a local, $Z(2)$ invariant potential. Integration
region for angle variables $\phi$ is restricted because
of Eq. (\ref{o3meas}). The usual $SO(3)$ measure can be recovered
if we set $\mu =3D 1$ and $V(\phi_{x}) =3D \ln (1-\cos^{2}\phi_{x})$.

This general phase structure described above is, in its main features,
valid for $d=3D4$ theory as well. The difference is that effective $d=3D2$
model for $3-d$ theory (\ref{18}) demonstrates a phase transition
in the absence of the invariant measure term ($\mu =3D 0$)
but there should not be a spontaneous breaking of $U(1)$ global symmetry
(this is the Kosterlitz-Thouless phase transition). The corresponding
model in $3-d$ theory could exhibit a spontaneous breaking of
global continuous symmetry if we measure it with the corresponding
source as described in this section. In the presence of the potential
term (\ref{19}) the phase transition disappears from the effective
theory (\ref{18}) in both dimensions.
The model defined in (\ref{16})-(\ref{18}) with (\ref{19}) we
call the $XY$ model with the local $Z(2)$ symmetry and use the notation
$XYL$ to distinguish it from the usual $XY$ model. Its study is the
subject of the next section.

\section{The $XYL$ model and the noncompact model in the weak coupling
region}

The $d$-dimensional $XYL$ model can formally be viewed as a combined model
of $XY$ models in the same dimension with ferromagnetic and
antiferromagnetic couplings.
Averaging is performed as a sum (difference) of the weights
$e^{\lambda S_l}$ of these models, $S$ is a density of the action, $l$ is a
link, if $l$ does not belong to $\Omega$ (if $l \in \Omega$).
Our particular case is slightly different, however, since the plaquettes in
adjacent time slices are not interacting, so that the only geometry of
2-surfaces is important. It is clear from (\ref{14})-(\ref{17}) that
we can define the corresponding partition function for 2-d theory as
\begin{equation}
Z_{\Omega} =3D  \prod_{t=3D1}^{N_t} \int \prod_{\bar{x}}
d \phi_{\bar{x}} e^{\mu \sum_x V(\phi_{\bar{x}})}
\prod_{l \ni L_t} \cosh (\lambda \cos \Delta \phi )
\prod_{l \in L_t} \sinh (\lambda \cos \Delta \phi ).
\label{20}
\end{equation}
\noindent   In our case,
$L_t \in \Omega$ is a set of closed loops on a time slice $t =3D const$ who=
se
geometry is connected with the closed surfaces in $d =3D 3$ dimensions and
\begin{equation}
\sum_{t=3D1}^{N_t} L_t =3D \mid \Omega_{p_0} \mid,
\label{21}
\end{equation}
\noindent where $\mid \Omega_{p_0} \mid$
is a number of time-like plaquettes on a surface $\Omega$.
Thus, as the first step in study of these systems we could digress of this
connection and investigate a pure $2-d$ theory considering $L_t$ as arbitra=
ry
closed loops. To do that we rewrite (\ref{20}) as
\begin{equation}
Z_{\Omega} =3D  \prod_{t=3D1}^{N_t} 2^{-N_l} \sum_{\sigma_l=3D\pm 1}
\prod_{l \in L_t} \sigma_l Z_{XYL}(\sigma),
\label{22}
\end{equation}
\noindent
\begin{equation}
Z_{XYL}(\sigma) =3D \int_{0}^{\pi} \prod_{x} d \phi_x
\exp [ \mu \sum_x V(\phi_x) +
\lambda \sum_{x,n} \sigma_n(x) \cos (\phi_x - \phi_{x+n}) ].
\label{23}
\end{equation}
\noindent
We omitted all bars here considering that $x =3D (x_1, x_2)$ to the end of =
this
section, $n =3D 1,2$. The symmetry properties of the model can be easily
deduced from the last two equations. At $\mu =3D 0$ the model is invariant
under global transformations
\begin{equation}
\phi_x \rightarrow \phi_x + const,
\label{24}
\end{equation}
\noindent
and under local discrette $Z(2)$ transformations
\begin{equation}
\phi_x \rightarrow \phi_x + \pi n_x, \
\sigma_k(x) \rightarrow - \sigma_k(x), \  k =3D \pm n,
\label{25}
\end{equation}
\noindent
where $n_x$ is an arbitrary integer. The last symmetry one sees directly
from (\ref{20}): $\cosh \lambda \cos \Delta \phi$ is identically invariant =
under
(\ref{25}) whereas $\sinh \lambda \cos \Delta \phi$ changes a sign if $n_x$ =
is
an odd number. Because the site variable $\phi_x$ must enter a closed loop =
even
number of times, the product of $\sinh \lambda \cos \Delta \phi$ is left in=
variant
under (\ref{25}). This is, of course, a consequences of the definition
(\ref{17}) where the variable $\phi$ is rather a link variable. Since
every link enters two or four times in a closed surface, the partition
function in (\ref{14}) is also invariant under the corresponding local
transformations. This property means that an expectation value of
any $Z(2)$ noninvariant quantity equals zero as a consequences of
Elitzur's theorem. However, the correlation functions which are invariant
may show a critical behaviour like in the $XY$ model showing the phase
transition of the Kosterlitz-Thouless type.
Following the definition (\ref{19}), the potential $V(\phi_x)$ breaks
the global symmetry (\ref{24}). Hence, the full theory is invariant
only under the discrete transformations (\ref{25}). We expect in this case
that the only phase is available in the whole range of couplings:
correlation functions are left finite at large distances that
corresponds to the perimeter law for the adjoint sources.

In a similar manner we can define the sine-Gordon model with the local
discrete symmetry. Choosing the potential (\ref{19}) in the sine-Gordon
form one gets
\begin{equation}
Z_{SG} =3D \int_{-\infty}^{\infty} \prod_x d\phi_x \sum_{k_{n}(x)}
\exp [ \mu \sum_x \cos 2\phi_x - \lambda \sum_{x,n}(\phi_x - \phi_{x+n} +
\pi k_{n}(x))^2 ].
\label{26}
\end{equation}
\noindent
The local symmetry $\phi_x \rightarrow \phi_x + \pi n_x$ presented here
is similar to the discrete symmetry (\ref{25}).
There is a spontaneous breaking of the global $Z(2)$ symmetry in the
standard sine-Gordon model. As in the previous case we expect that the
only phase with unbroken local $Z(2)$ symmetry can be found in the model
(\ref{26}) and all the properties of the model are very close to those
of $XYL$ model in the region of weak coupling.

A dual representation for $XYL$ model one obtains similarly to $XY$ model.
Summing over $\sigma_n(x)$ we have on the dual lattice,
keeping the notations of the original lattice for the indices
\begin{eqnarray}
\sum_{\sigma_l=3D\pm 1}\prod_{l \in L_t} \sigma_l Z_{XYL}(\sigma) =3D
\sum_{s_x =3D -\infty}^{\infty} \sum_{j_n(x) =3D -\infty}^{\infty}
\prod_{l \ni L} I_{2(s_x - s_{x+n} - j_n(x))} (\lambda )  \nonumber  \\
\prod_{l \in L} I_{2(s_x - s_{x+n} - j_n(x)) + 1} (\lambda )
\prod_p B_{2j_p}(\mu ),
\label{27}
\end{eqnarray}
\noindent
$I_k$ is the modified Bessel function. The plaquette variable $j_p$ is
defined as $j_p =3D j_n(x) + j_m(x+n) - j_n(x+m) - j_m(x)$. The coefficients
$B_m$ are the following Fourier components
\begin{equation}
e^{\mu V(\phi)} =3D \sum_m B_{2m}(\mu ) e^{2im\phi}.
\label{28}
\end{equation}
\noindent
Only even terms are present here  because of the property
\begin{equation}
V(\phi + \pi n) =3D V(\phi).
\label{29}
\end{equation}
$L$ in (\ref{27}) is a set of links dual to links $L_t$ forming
a closed loop on the original lattice.

To display the expected phase structure two types of correlation
functions can be defined and evaluated in the $XYL$ model.
The first one is the spin-spin correlation function
\begin{equation}
\Gamma_{m,k} =3D < e^{im \phi_0} e^{-ik \phi_R} >.
\label{30}
\end{equation}
\noindent
Let us discuss its behaviour when $R \rightarrow \infty$ at
asymptotically small and large values of $\lambda$. We omit all
calculations as they follow step by step similar calculations
in $XY$ model. First of all it is clear that $\Gamma_{m,k} =3D 0$
if $m$ and/or $k$ is an odd number in force of the symmetry (\ref{25}).
One sees this from the calculation of the dual representation (\ref{27})
or from the mean-field approach on the original lattice even before
the summation over closed loops. The behaviour of the correlation functions
$\Gamma_{2m,2k}$ is very close to spin-spin correlations in $XY$ model.
In the next section we shall argue that these correlators carry a memory
of the adjoint Wilson loop of the original model (\ref{1}). Therefore,
we expect that closed loops $L_t$ do not influence essentially its behaviour
since they have appeared after summation over $Z(2)$ variables and the
adjoint sources do not feel these configurations. Neglecting all $L_t$
one finds at $\mu =3D 0$ the following asymptotics ($m=3Dk$)
\begin{equation}
\Gamma_{2m} \approx \exp [-R \ln \frac{(2m)!}{\lambda^{2m}}], \
\lambda \rightarrow 0,
\label{31}
\end{equation}
\noindent
\begin{equation}
\Gamma_{2m} \approx \exp [-\frac{2m^2}{\lambda} \ln R], \
\lambda \rightarrow \infty .
\label{32}
\end{equation}
\noindent
These asymptotics exhibit the Kosterlitz-Thouless phase transition: the
expectation value of the spin $< e^{2im \phi_x} > =3D 0$ and the global
symmetry (\ref{24}) is unbroken at all $\lambda$. The potential $V(\phi)$
destroys the global symmetry: one finds that $< e^{2im \phi_x} >$ differs
from zero and the correlation function $\Gamma_{2m}$ is finite in the limit
$R \rightarrow \infty$ at all values of $\lambda$. This may correspond
to the perimeter law behaviour of the adjoint loops as we have discussed
before.

The second type of correlation functions is defined as follows
\begin{equation}
\Gamma_{\sigma_c} =3D < \prod_{l \in C} \sigma_l >,
\label{33}
\end{equation}
\noindent
where $C$ is a closed loop and
\begin{equation}
\Gamma_{\sigma_c}^W =3D < e^{i\phi_0} \prod_{l \in L} \sigma_l e^{-i\phi_R}=
>,
\label{34}
\end{equation}
\noindent
where $L$ is a path between points 0 and $R$.
Both correlation functions are invariant under local transformations
(\ref{25}). The correlation function (\ref{33}) defines a time-like
plaquette dependence of the expectation value of $Z(2)$ Wilson loop
(see next section for detail) while (\ref{34}) defines a similar dependence
of $U(1)$ Wilson loop which could also be determined in the model (\ref{9}).
Unlike the spin-spin correlation functions, both
$\Gamma_{\sigma_c}$ and $\Gamma_{\sigma_c}^W$ decrease exponentially when
$C \rightarrow \infty$ and $R \rightarrow \infty$, correspondingly, at all
values of $\lambda$. One finds this from the simple estimates
on the original lattice when $\lambda$ tends zero and on the dual lattice
in the opposite case. This is a hint on the area law for $Z(2)$ Wilson loop
at all couplings and we shall verify this in the next
section. All this is valid only in the case of absence of the loops $L_t$.
However, the closed loops $L_t$ might be of great importance here, especial=
ly
in the region of small gauge coupling $\lambda \rightarrow \infty$
and this influence must be studied separately.

We are ready now to present an effective noncompact model for $A_0$ gauge
potential in the weak coupling limit. Our result is the noncompact
sine-Gordon model defined in (\ref{26}) with modification that includes
the summation over 2-surfaces. The following formulae and the finite
result are valid for the theory in $d$-dimensions. The usual way of the
calculation in the region $\lambda \rightarrow \infty$ is to expand
$\cos \Delta \phi \approx 1-\frac{1}{2}(\Delta \phi)^2$. However,
this implies a loss of the local symmetry (\ref{25}). To recover the
symmetry we rewrite the expression (\ref{20}), omitting an irrelevant
here product over $t$, as
\begin{equation}
Z_{\Omega} =3D  \int \prod_x d \phi_x e^{\mu \sum_x V(\phi_x)}
\sum_{r_l =3D 0,1}  \exp [\lambda \cos (\Delta \phi_x + \pi r_l)]
\prod_{l \in L_t} (-1)^{r_l}.
\label{35}
\end{equation}
\noindent
Up to infinite constant we may sum in (\ref{35}) over $r_l$ from $-\infty$
to $+\infty$. This infinite constant will be cancelled from all expectation
values. After this we can expand $\cos (\Delta \phi_x + \pi r_l)$ around
unity. This leads us to
\begin{equation}
Z_{\Omega} =3D  \int_{-\infty}^{\infty} \prod_x d \phi_x
e^{\mu \sum_x V(\phi_x)} \sum_{r_l =3D -\infty}^{\infty}
\exp [ - \frac{\lambda}{2}  \sum_l (\Delta \phi_x + \pi r_l)^2]
\prod_{l \in L_t} (-1)^{r_l}.
\label{36}
\end{equation}
\noindent
This procedure can be made more rigorous in the following way.
Integrating out $\phi_x$-variables we get
\begin{equation}
Z_{\Omega} =3D \sum_{k_l} \sum_{m_x} \prod_{l \ni L_t} I_{2k_l}(\lambda)
\prod_{l \in L_t} I_{2k_l + 1}(\lambda)
\prod_x B_{2m_x}(\mu) \delta (m_x + \sum_{n=3D-d}^{d}k_n(x)).
\label{37}
\end{equation}
\noindent
$B_m$ was defined in (\ref{28}), the symbol $\delta$ means
the Kronecker delta function and $k_{-n}(x)=3D-k_n(x)$. Using the Poisson
summation formula to calculate the sum over $k_l$ we may change
the Kronecker delta into the Dirac delta function since $k_n (x)$
becomes a continuous variable. This again generates an infinite constant
which is cancelled from locally invariant expectation values.
On the other hand this preserves the local periodicity of the original
action. Taking asymptotic of the Bessel function at large $\lambda$
\begin{equation}
I_k (y) \approx \frac{1}{\sqrt{2\pi y}}e^{y} \exp [-\frac{k^2}{2y}],
\label{38}
\end{equation}
\noindent
we can integrate over $k_n(x)$ and sum over $m_x$.
This yields the formula (\ref{36}) justifying the simple calculations
presented above. Substituting (\ref{36}) into (\ref{14}) and remembering
that in our case the closed loops $L_t$ in the plane $t=3Dconst$ are connec=
ted
to the 2-surfaces we get the final expression for the noncompact model
\begin{eqnarray}
Z =3D e^{-\lambda N_p} (\cosh \lambda )^{N_{\bar{p}}} 2^{N_l}
\sum_{\Omega} (\tanh \lambda )^{\mid \Omega_{\bar{p}} \mid}  \nonumber	\\
\prod_{t=3D1}^{N_t} \int_{-\infty}^{\infty} \prod_x d \phi_x
e^{\mu \sum_x V(\phi_x)} \sum_{r_{p_0} =3D -\infty}^{\infty}
\exp [ - \frac{\lambda}{2}  \sum_{p_0} (\Delta \phi_x + \pi r_{p_0})^2]
\prod_{p_0 \in \Omega} (-1)^{r_{p_0}}.
\label{39}
\end{eqnarray}
\noindent
Here, $r_{p_0}$ belongs to a time-like plaquette and the effective action
is diagonal in $t$, so that we consider $\phi_x$ as the site variable.
(\ref{39}) is a noncompact analogy of the compact model (\ref{9}) in the
weak coupling region. The model is clearly different both from the naive
noncompact model and from the model proposed in \cite{pol} to include
the local $Z(2)$ symmetry and the sum over 2-surfaces as the result
of the summation over all $Z(2)$ configurations of the original Wilson mode=
l.

\section{Wilson loops in the effective models and the area law}

The usual way to find out the confining properties of the pure gauge theory
is to study the behaviour of the fundamental Wilson loop which gives
a potential between the static quark-antiquark pair. Our approach allows us
to restrict the calculations to the $Z(2)$ Wilson loop
\begin{equation}
W_z (C) =3D \prod_{l \in C} z_l,
\label{40}
\end{equation}
\noindent
where $C$ is a closed restangular loop in the $t-x$ plane. The expectation
value of the loop is calculated with the partition function (\ref{9})
and can be written down by help of the representation (\ref{13}) in the form
\begin{eqnarray}
<W_z (C)> =3D Z^{-1}\sum_{z_{\mu}(x)} \int \prod_{x} D\bar{U}_{0}(x)
\prod_{\bar{p}} \left [ \cosh \lambda + \sinh \lambda  Z(\partial \bar{p})
\right ]   \nonumber   \\
\prod_{p_{0}} \left [ \cosh (\lambda \cos \Delta \phi )
+  \sinh (\lambda \cos \Delta \phi) Z(\partial p_0) \right ]
\prod_{l \in C} z_l.
\label{41}
\end{eqnarray}
\noindent
Let $W$ be the $SU(2)$ Wilson loop $W=3D\frac{1}{2} Tr \prod_{l \in C} U_l$.
=46rom obvious inequality $<W> \leq <W_z>$ it follows that if $Z(2)$ Wilson
loop obeys the area law behaviour, $SU(2)$ Wilson loop will show the same
feature in the theory defined by (\ref{9}) with the measure (\ref{19}).
Presumably, even stronger statement can be formulated if we remember
the result of \cite{seiler1}, namely
$$
<W>_{SU(N)} \  \leq  \ <W_z>_{Z(N)},
$$
which tells us that $SU(N)$ Wilson loop will obey the area law in the Wilson
$SU(N)$ gauge model if $Z(N)$ Wilson loop in $Z(N)$ gauge model obeys
area law at the coupling constant $g^2/N$. As we have discussed many times,
the main effect of $\cos \Delta \phi$ in (\ref{9}) is to disorder the system
at large $\lambda$. Thus, we expect that the following inequality
has to be fulfilled
\begin{equation}
<W>_{SU(2)} \  \leq \  <W_z>,
\label{42}
\end{equation}
\noindent
where $<W_z>$ is calculated in the statistical ensemble defined by
the partition function (\ref{9}).

First of all we need to calculate representations for $W_z$ both in
the compact model (\ref{14}) and in the noncompact model (\ref{39}).
Summing over $z_{\mu}(x)$ in (\ref{41}) one obtains
\begin{eqnarray}
<W_z (C)> =3D Z^{-1}(\cosh \lambda)^{N_{\bar{p}}} 2^{N_l}
\sum_{S(\partial C)} (\tanh \lambda)^{\mid S_{\bar{p}} \mid}
\sum_{\Omega / p\in S(\partial C)}
(\tanh \lambda )^{\mid \Omega_{\bar{p}} \mid}
\nonumber   \\
\sum_{\sigma_{p_0}=3D\pm 1} \int_{-\pi /2}^{\pi /2} \prod_{x} d \phi_x
\exp [ \mu \sum_x V(\phi_x) +
\lambda \sum_{p_0} \sigma_{p_0} \cos (\Delta \phi) ]
(\prod_{p_0 \in S(\partial C)}\sigma_{p_0})(\prod_{p_0 \in \Omega}
\sigma_{p_0}).
\label{43}
\end{eqnarray}
\noindent
Interpretation of this formula is obvious: the expectation value of the
$Z(2)$ Wilson loop is expressed through the sum over all possible
surfaces $S$ on the lattice whose boundary is the loop $C$,
$\mid S_{\bar{p}} \mid$ is a number of space-like plaquettes on the surface
$S$. Every term in the sum over $S$ includes the sum over all 2-surfaces
$\Omega$ which have no common plaquettes with given surface $S$. Repeating
all the steps from the end of the previous section we come to the expression
for the Wilson loop at large $\lambda$ of the form (up to a constant)
\begin{eqnarray}
<W_z (C)> =3D Z^{-1}(\cosh \lambda)^{N_{\bar{p}}} 2^{N_l}
\sum_{S(\partial C)} (\tanh \lambda)^{\mid S_{\bar{p}} \mid}
\sum_{\Omega / p\in S(\partial C)}
(\tanh \lambda )^{\mid \Omega_{\bar{p}} \mid}
\sum_{r_{p_0}=3D-\infty}^{\infty}   \nonumber   \\
\int \prod_{x} d \phi_x
\exp [ \mu \sum_x V(\phi_x) -
\frac{\lambda}{2} \sum_{p_0} (\Delta \phi + \pi r_{p_0})^2 ]
(\prod_{p_0 \in S(\partial C)} (-1)^{r_{p_0}})(\prod_{p_0 \in \Omega}
(-1)^{r_{p_0}}).
\label{44}
\end{eqnarray}
\indent

The treatment of the adjoint Wilson loop $W_{ad}(C)$ in the present approach
is much simpler since it does not include $Z(2)$ variables. We may put down
the following equation for the adjoint loop making use of the diagonality of
the effective action in time indices and the fact that the global symmetry
$\phi \rightarrow -\phi$ is unbroken in our model
\begin{equation}
<W_{ad}(C)> =3D  <\prod_t e^{2i(\phi_0 (t) - \phi_R (t))}>
\label{45}
\end{equation}
\noindent
(we recall that in our approach the space gauge potentials $A_n =3D 0$).

To understand the behaviour of expectation values in Eqs.
(\ref{43})-(\ref{45}) we consider first the approximation where
only time-like plaquettes are taken into account. We have
from Eq. (\ref{43}) in this case
\begin{eqnarray}
<W_z (C)>_0 =3D Z^{-1} \sum_{\sigma_{p_0}=3D\pm 1}
(\prod_{p_0 \in S_{min}(\partial C)}\sigma_{p_0}) \nonumber   \\
\int \prod_{x} d \phi_x \exp [ \mu \sum_x V(\phi_x) +
\lambda \sum_{p_0} \sigma_{p_0} \cos (\Delta \phi) ],
\label{46}
\end{eqnarray}
\noindent
where $S_{min}$ is the minimal surface enclosed by the loop $C$
and lying in the $(t-x)$-plane (only such a surface survives in this limit
since to go out of the $(t-x)$-plane at least four space-like plaquettes
must be present). Thus, we come to the expression for the fundamental
Wilson loop of the form (summing up over $\sigma$)
\begin{equation}
<W_{z}(C)>_0 =3D <\prod_{p_0 \in S_{min}} \tanh (\lambda \cos \Delta \phi )=
>,
\label{47}
\end{equation}
\noindent
which allows us to make a simple estimate of the module of the expectation
value
\begin{equation}
<\mid W_{z}(C) \mid>_0 \  \leq \  (\tanh \lambda)^{S_{min}}.
\label{48}
\end{equation}
\noindent
Similar appraisals performed for noncompact formulation (\ref{44})
lead to the result
\begin{equation}
<W_{z}(C)>_0 =3D <\prod_{p_0 \in S_{min}}
\frac{\theta_4 (\gamma , i\lambda \pi \Delta \phi /2)}
{\theta_3 (\gamma , i\lambda \pi \Delta \phi /2)}>.
\label{49}
\end{equation}
\noindent
Here, $\theta_i$ is the Jacobi theta-function and we have denoted
$\gamma =3D \exp(-\frac{\lambda}{2}\pi^2)$. From here, we find the following
bound in the region $\gamma \rightarrow 0$ ($g^2 \sim 0$)
\begin{equation}
<\mid W_{z}(C) \mid>_0	\ \leq \  e^{-4\gamma S_{min}}.
\label{50}
\end{equation}
\indent
The adjoint Wilson loop (\ref{45}) in the present approximation reduces to
the form (\ref{30}) displaying, hence, the expected perimeter law
behaviour (see (\ref{31}), (\ref{32}) and the discussion around them).

One can say, of course, that, in fact, the time-like plaquette
approximation is close to the strong coupling regime where only
$S_{min}$ survives in the thermodynamical limit. As such, this is true
and the bound (\ref{48}) cannot be trusted in the weak coupling region.
However, we consider that the bound (\ref{50}) is a great achievement
of the whole approach. We would like to recall at this point that
naive noncompact lattice theory with time-like plaquettes only
does not show area law behaviour at any couplings \cite{ncmpt1}.
Thus, the periodicity of the compact theory, which we have tried to
reproduce in our noncompact effective model, can be of great importance
for confinement at any couplings (actually, the noncompact theory
(\ref{39}) is valid only at small couplings with respect to the original
model (\ref{9}) but, as such, can be formally studied at all couplings).
We find it suprisingly good that the bound (\ref{50}) is in agreement
with the expected asymptotic freedom behaviour of the string tension
that is
\begin{equation}
\alpha =3D 4\gamma =3D 4 \exp[-\frac{\pi^2}{g^2}].
\label{51}
\end{equation}
\noindent
Eq.(\ref{51}) leads to the usual perturbative relation
\begin{equation}
g^2 =3D - \frac{\pi^2}{2 \log (a \Lambda)},
\label{52}
\end{equation}
\noindent
if we impose the condition $\frac{d\alpha}{da}=3D0$, where $a$ is
the lattice spacing.

Calculations in the full theory with space-like plaquettes included
demand a very complicated treatment of the summation over 2-surfaces
and will be considered in a separate publication \cite{sur}.
Let us now discuss qualitatively what we expect in this case and why we
think the theory (\ref{9}) could be confining in the weak coupling region.
We turn first to the compact theory and to
the corresponding representation for the Wilson loop (\ref{43}).

The expectation value of the Wilson loop in ordinary
$Z(N)$ gauge theory can be written down using (\ref{43}) as
\begin{equation}
<W_z (C)> =3D Z^{-1}(\cosh \lambda)^{N_p} 2^{N_l}
\sum_{S(\partial C)} (\tanh \lambda)^{\mid S \mid}
\sum_{\Omega / p\in S(\partial C)} (\tanh \lambda )^{\mid \Omega \mid}.
\label{53}
\end{equation}
\noindent
In the strong coupling region ($\tanh \lambda \sim 0$) only surfaces
of small sizes contribute to the partition function and we are allowed
to neglect the restriction in the numerator of the last formula which
forbids the summation over surfaces $\Omega$ if they have common
plaquettes with given surface $S(\partial C)$. Hence, in this region
we approximately have
\begin{equation}
<W_z (C)> \sim \sum_{S(\partial C)} (\tanh \lambda)^{\mid S \mid} L(S),
\label{54}
\end{equation}
\noindent
where $L(S)$ is a number of the surfaces spanning the loop $C$. Below
the critical point this sum is known to be convergent and we have in the
thermodynamical limit the area law
$$
<W_z (C)> \sim (\tanh \lambda)^{S_{min}}.
$$
As $\lambda$ grows, however, $L(S)$ becomes growing faster than
$(\tanh \lambda)^{S}$ goes to zero. This leads to the phase transition
and to the perimeter law for the Wilson loop.

Two factors may, hopefully, change this behaviour. The first of them is
already contained in the pure $Z(N)$ system. When we move towards
the critical point, Eq.(\ref{54}) ceases to be a proper equation for
the expectation value even below $\lambda_{cr}$. The reason is
that at the same time the number of 2-closed surfaces contributing to
$<W_z (C)>$ is much increased, too, and  we may not neglect the restriction
forbidding the summation over $S(\partial C)$ having common plaquettes
with closed surfaces. This strongly reduces effective $L(S)$
(or, equivalently, lowers the effective coupling) but not sufficiently
to preserve the system of the transition to the deconfinement phase.
The second, most essential factor, appears from the integration
over $SU(2)/Z(2)$ subgroup on the time-like plaquettes. To see that we
consider the following analog of Eq.(\ref{54}) which we expect to
approximately hold for $<W_z (C)>$ in the whole range of couplings
for Eq.(\ref{43})
\begin{equation}
<W_z (C)> \approx \sum_{S} (\tanh \lambda)^{\mid S_{\bar{p}} \mid}
L_{eff}(S) < \prod_{p_0 \in S} \sigma_{p_0} >.
\label{55}
\end{equation}
\noindent
Here, $S=3DS(\partial C)$, $L_{eff}(S)$ is the effective number of surfaces
containing $\mid S \mid$ plaquettes reduced by the restriction stressed
just above (the first factor). $<...>$ is calculated in the statistical
ensemble defined by the partition function (\ref{14}) either with the
ordinary $SU(2)$ measure or with the measure (\ref{19}). On every time
slice $< \prod_{p_0 \in S} \sigma_{p_0} >$ forms a set of closed loops
where this expectation value is reduced to the correlation function
$\Gamma_{\sigma_c}$ (\ref{33}) in $XYL$ model. This correlation function
is expected to decrease exponentially at all values of coupling constant,
which implies the area law behaviour for the expectation value in (\ref{55}=
).
We introduce now the effective coupling for time-like plaquettes as
\begin{equation}
< \prod_{p_0 \in S} \sigma_{p_0} > =3D [\gamma_{eff}(\lambda)]^{S_0},
\label{56}
\end{equation}
\noindent
where $S_0$ is a number of time-like plaquettes on the given surface
$S(\partial C)$. The simplest estimates show that $\gamma_{eff}$ decreases
much faster than the corresponding coupling of the pure $Z(N)$ gauge model
where $\tanh \lambda \rightarrow 1$ when the gauge coupling goes to zero.
We suppose now that $L_{eff}(S)$ including a number of time-like plaquettes
$S_0 > S_{min}$ is suppresed by the effective coupling $\gamma^{S_0}$.
Then, we have
\begin{equation}
<W_z (C)> \approx (\gamma_{eff})^{S_{min}}
\sum_{S} (\tanh \lambda)^{\mid S_{\bar{p}} \mid}
L_{eff}(S) \delta (S_0 - S_{min}).
\label{57}
\end{equation}
\noindent
This formula produces the expected area law in the thermodynamical limit
and when $S_{min} \rightarrow \infty$ for all couplings.
Yet, the critical point may still exist since the number of the surfaces
with fixed $S_0$ but different configurations of space-like plaquettes
can be larger than $(\tanh \lambda)^{\mid S_{\bar{p}} \mid}$. The point
is that in this critical point the string tension does not turn into
zero as could be seen from the last equation unlike $Z(N)$ gauge model.
Hence, we may interpret the critical point (if it really exists) as a
point of a crossover transition (or a roughening transition) which
has been observed in Monte-Carlo simulations many times and which
does not affect the string tension.

Extending now these observations to the noncompact representation
(\ref{44}) we observe that the only difference is in $\gamma_{eff}$
which becomes in this case
\begin{equation}
(\gamma_{eff})^{S_0} =3D < \prod_{p_0 \in S} (-1)^{r_{p_0}} >.
\label{58}
\end{equation}
\noindent
Now, $<...>$ is calculated with the partition function (\ref{39}).
We have the same qualitative feature of the decreasing of an effective
coupling as in the previous case. Hence, all discussed arguments can be
applied for the noncompact representation. As a result, we are convinced
that both the model (\ref{9}) and the noncompact model (\ref{39})
could be confining models with the same mechanism of confinement.
These arguments have a general character, therefore we belive that
discussed mechanism could be a common mechanism for $SU(N)$ both in
$d=3D3$ and in $d=3D4$ dimensions which preserves the system from the
deconfining transition.
For the real proof, however, one has to prove that representation (\ref{55})
is a reliable one in the weak coupling region and to find a bound
on $L_{eff}(S)$ at large $S$ \cite{sur}.

\section{Summary}

In this article we have assumed that the compact Wilson theory and noncompa=
ct
lattice Yang-Mills theory belong to different universality classes and
the latter in its naive form is unable to confine. We have demonstrated that
if we take into account $Z(N)$ local invariance of the Wilson compact model
and find weak coupling limit of the model after summation over $Z(N)$
variables the resulting model differs from the naive lattice Yang-Mills
theory and possesses the necessary confining properties. Our real
calculations were done for the $SU(2)$ Wilson model in three dimensions
but it is clear from the whole procedure that the generalization
to $SU(3)$ gauge group and to higher dimensions is straightforward.

Let us turn once more to our starting model (\ref{9}) and summarize
the facts justifying its investigation. The lines of arguments, leading
from (\ref{1}) to (\ref{9}), indicate that the partition function
(\ref{9}) should describe the main contribution to the original partition
function in the gauge where all $SU(2)/Z(2)$ space-gauge matrices
are chosen to be as close to unity as possible. Truly, we have not given
a rigorous proof of that but hopefully it can be done using a method
developed in \cite{zw1}. Qualitative discussion of what might happen
if we have taken into account corrections resulting from integration over
noncompact Yang-Mills space-gauge potentials shows that these corrections
cannot essentially spoil or even change the effective action in (\ref{9})
(we speak of corrections to $Z(2)$ configurations; certainly,
there will be a lot of contributions not directly related to $Z(2)$).
Also, the inequality (\ref{42}) is to be emphasized at this point.
Had it been proved rigorously it would alone justify the investigation
of the model (\ref{9}) because our hope is to prove an area law
for the space-time Wilson loop in this model for weak coupling.
Having (\ref{42}) in hand, one can claim the same is valid for full
$SU(2)$ theory.

Summarizing our results we would like to mention two points.
The first of them are local discrete transformations (\ref{25}).
It was essential to preserve this symmetry in the noncompact theory.
Let us suppose that we neglect the symmetry, for example, expanding
$\cos \Delta \phi$ simply as $1-(\Delta \phi)^2 /2$. It is easy to verify
then that we would get almost no effect on the coupling $\gamma_{eff}$.
Why it is so can be understood if we turn to the expression (\ref{9})
for the partition function. Substituting this expansion into time-like
plaquettes we come to the Yoneya expansion which we already discussed.
In our terms it would lower somehow the effective coupling $\gamma_{eff}$
but obviously not sufficient to shift the critical point to zero.
As shown by our estimations including the invariant measure
cannot save the situation. We would, thus, expect that the similar
symmetry must be present in a full noncompact version of the Wilson
gauge theory with the space gauge-potential included.

The second point concerns again the invariant measure. In the end of
section 2 and in section 3 we have discussed the role of the
measure in providing the adjoint Wilson loop with the perimeter
law for large loops at any couplings. Now, what can we say about
the contribution of the measure into the fundamental ($Z(2)$ in our case)
Wilson loop? The bounds (\ref{48}), (\ref{49}) do not depend
on whether the measure is present in the partition function or not.
Thus, in the strong coupling region the very fact of confinement
does not depend on the invariant measure. That is clear because
confining forces in this region have origin in the very
$Z(N)$ system and the $SU(2)/Z(2)$ part of the action is not important
here for $Z(2)$ Wilson loop. However, the situation could be quite different
in the weak coupling region or if we consider $SU(2)$ Wilson loop.
This is because the term $\mu V(\phi)$ affects significantly the expectation
value in (\ref{55}) or in (\ref{58}) at large $\lambda$ lowering
the effective constant $\gamma_{eff}$. The question may arise at this point,
namely how to calculate the coupling $\mu$ at the invariant potential.
As far as our calculations are concerned, we can work with the ordinary
$SU(2)$ measure since all our Gaussian integrals are well defined
in this case, too. It could be possible then to define $\mu$
in a renormalizable theory. Let us imagine that we have managed to
calculate an expectation value of the Wilson loop and to find the string
tension to be a function of $g^2$ and $\mu$. Then we can calculate the
lines of constant physics which will give us the condition on the
coupling constants as a function of the lattice spacing.
Thus, $\mu =3D \mu (a)$ we will find under assumption that the expectation
value of the Wilson loop in the theory with the term $\mu V$ is on the
same line of constant physics as the Wilson loop in the theory
with the usual $SU(2)$ measure. In this way the invariant term
$\mu V$ could be included into the renormalizable noncompact Yang-Mills
theory.

Concluding the summary we want to make some remarks on
the intriguing question what mechanism of confinement
do our approach and calculations support?  It may seem, at least at
the first sight, that this approach is more in the spirit of the
mechanism of confinement caused by the $Z(N)$ vortex condensate
proposed in \cite{mack}. It is plausible that it is the case but in order
to show this the condensate (existence of which is a sufficient condition
for confinement) defined in \cite{mack} has to be calculated in the
weak coupling region. Since this problem is much easier to solve
in our approach, we will return to this point in the next publication.
At the present stage, however, we do not think that our approach
is in contradiction with the monopole mechanism of confinement, e.g.
the one discussed in the abelian projection method. Indeed, our
effective 2-dimensional (the same can be said of $3-d$ theory) $XYL$
model includes vortex configurations (not that ones mentioned just above)
playing an important role in the phase structure discussed in great
details in sections 2 and 3. Truly, these properties concerned
mainly the behaviour of the adjoint Wilson loop but it is possible
that these configurations become relevant to the fundamental loop at the
weak coupling. These vortices might be reminiscent of monopole
configurations, contained in the $SU(2)/Z(2)$ sector of the original
three (or four) dimensional model. Defining $U(1)$ Wilson loop
in the model (\ref{9}) by help of (\ref{34}) one investigates this
possibility within presented approach.
\\
\\
\\
The work was partially supported by NATO Linkage Grant No.930224
and by Grant of the International Science Foundation No.K4W 100.


\begin{thebibliography}{99}

\bibitem{ncmpt1} O.A.~Borisenko, V.K.~Petrov, G.M.~Zinovjev and
J.~Boh\'a\v cik,
Phase structure and confinement properties of noncompact gauge theories I,
hep-lat/9508001, 1995.
\bibitem{pol} K.~Johnson, L.~Lelouch, J.~Polonyi, Nucl.Phys. B367 (1991) 67=
5.
\bibitem{gopfert} M.~G\'{o}pfert, G.~Mack, Com.Math.Phys. 81 (1981) 97;
82 (1982) 545.
\bibitem{mack} G.~Mack, V.~Petkova, Ann.Phys. 125 (1980) 117.
\bibitem{mack2} G.~Mack, Phys.Lett. B78 (1978) 263.
\bibitem{su2gl} O.A.~Borisenko, V.K.~Petrov, G.M.~Zinovjev,
Theor.Mat.Fiz. 80 (1989) 381.
\bibitem{l94} O.A.~Borisenko, V.K.~Petrov, G.M.~Zinovjev,
Nucl.Phys. B (Proc. Suppl.) 42 (1995) 466.
\bibitem{wilson} K.~Wilson, Phys.Rev. D10 (1974) 2445.
\bibitem{zw1} D.~Zwanziger, Nucl.Phys. B378 (1992) 525.
\bibitem{zw2} D.~Zwanziger, Nucl.Phys. B412 (1994) 657.
\bibitem{yon} T.~Yoneya, Nucl.Phys. B144 (1978) 195.
\bibitem{seiler} K.~Osterwalder, E.~Seiler, Ann. Phys. 110 (1978) 440.
\bibitem{seiler1} E.~Seiler, Gauge Theories as a Problem of
Constructive Quantum Field Theory and Statistical Mechanics,
Springer-Verlag Berlin Heidelberg New-York, 1982.
\bibitem{sur} Surfaces in a modified $Z(N)$ model and the string
tension in the weak coupling region, in preparation.


\end{thebibliography}
\end{document}